\documentclass[aps,prapplied, twocolumn, reprint,superscriptaddress,showpac,longbibliography]{revtex4-1}

\usepackage[colorlinks=true,citecolor=blue,linkcolor=magenta]{hyperref}
\hypersetup{
pdfauthor = {},
colorlinks = true, linkcolor = blue, urlcolor=blue, bookmarksnumbered =  true}

\usepackage{units}
\usepackage{graphicx}
\usepackage{amsmath}
\usepackage{physics}

\begin{document}

\title{Real-space probing of the local magnetic response of thin-film superconductors using single spin magnetometry}
\author{D. Rohner}
\altaffiliation{These authors contributed equally to this work}
\affiliation{Department of Physics, University of Basel, Klingelbergstrasse 82, Basel CH-4056, Switzerland}
\author{L. Thiel}
\altaffiliation{These authors contributed equally to this work}
\affiliation{Department of Physics, University of Basel, Klingelbergstrasse 82, Basel CH-4056, Switzerland}
\author{B. M\"uller}
\affiliation{Physikalisches Institut and Center for Quantum Science (CQ) in LISA$^+$, Universit\"at T\"ubingen, Auf der Morgenstelle 14, D-72076 T\"ubingen, Germany}
\author{M. Kasperczyk}
\affiliation{Department of Physics, University of Basel, Klingelbergstrasse 82, Basel CH-4056, Switzerland}
\author{R. Kleiner}
\affiliation{Physikalisches Institut and Center for Quantum Science (CQ) in LISA$^+$, Universit\"at T\"ubingen, Auf der Morgenstelle 14, D-72076 T\"ubingen, Germany}
\author{D. Koelle}
\affiliation{Physikalisches Institut and Center for Quantum Science (CQ) in LISA$^+$, Universit\"at T\"ubingen, Auf der Morgenstelle 14, D-72076 T\"ubingen, Germany}
\author{P. Maletinsky}
\email{patrick.maletinsky@unibas.ch}
\affiliation{Department of Physics, University of Basel, Klingelbergstrasse 82, Basel CH-4056, Switzerland}

\pacs{78.67.-n, 84.40.Ba, 42.50.Ct, 81.05.ug}

\date{\today}

\begin{abstract}

We report on direct, real-space imaging of the stray magnetic field above a micro-scale disc of a thin film of the high-temperature superconductor YBa$_2$Cu$_3$O$_{7-\delta}$ (YBCO) using scanning single spin magnetometry. Our experiments yield a direct measurement of the sample's local London penetration depth and allow for a quantitative reconstruction of the supercurrents flowing in the sample as a result of Meissner screening. These results show the potential of scanning single spin magnetometry for studies of the nanoscale magnetic properties of thin-film superconductors, which could be readily extended to elevated temperatures or magnetic fields. 

\end{abstract}

\maketitle

\section{Introduction}

Thin-film superconductors are of scientific interest and ever-increasing technological importance. 
For example, such thin films offer possibilities to systematically explore fundamental properties of cuprate superconductors\,\cite{Bovzovic2016} or allow for exquisite tunability of fundamental properties such as the superconductors critical temperature\,\cite{Caviglia2008}.
In the emerging field of quantum technologies, thin-film superconductors form the basis for superconducting quantum circuits, such as quantum bits or low-loss, microwave resonators\,\cite{Wendin2017,Muller2017} and they are also becoming increasingly important in single-photon detection\,\cite{Dauler2014}. These developments and especially the increased technological relevance of micro- and nano-structured superconductors call for novel tools to locally probe superconductivity in such materials with high sensitivity and nanoscale resolution. 

Magnetic probes are particularly well suited for this task, as they offer direct access to the local magnetic susceptibility and thereby a defining feature, i.e. perfect diamagnetism, of the superconducting state. Indeed, various such probes, including magnetic force microscopy (MFM)\,\cite{Badia1998, Kim2012,Luan2010}, scanning Hall-probes\,\cite{Chang1992, Brisbois2014} and superconducting quantum interference device (SQUID)\,\cite{Lin2015,Embon2017,Vasyukov2013} magnetometry have been developed and successfully employed for this purpose. Despite the remarkable insight, such experiments offered into microscopic properties of superconductors, they still suffer from their high invasiveness (MFM), reduced spatial resolution (Hall probes) or the limited range of temperature and magnetic field, in which they operate (SQUID). These limitations warrant the exploration of novel approaches to probe the local magnetic response of superconductors, which would allow, e.g. for studying them close to or above their critical temperature with nanoscale resolution.  

In this work, we introduce a possible solution to this bottleneck by employing scanning single spin magnetometry using a Nitrogen-Vacancy (NV) electronic spin in diamond\,\cite{Rondin2014} for magnetic probing of a micro-structured thin-film superconductor.
We demonstrate nanoscale imaging of the Meissner screening fields above a superconducting disc and employ a quantitative model to extract the local value of the London penetration depth $\lambda$ in our sample. 
Furthermore, we employ reverse propagation of the experimentally measured magnetic field map to reconstruct the screening supercurrents circulating in the sample and thereby address spatial inhomogeneities in the superconducting state. 
Our findings complement recent advances in using ensemble NV magnetometry to study Meissner screening or vortices in superconductors\,\cite{Waxman2014, Alfasi2016, Nusran2018,Bouchard2011} and promote this promising sensing technology to applications on the nanoscale.
While we have previously reported on nanoscale studies of vortices in type-II superconductors\,\cite{Thiel2016}, we here focus on the magnetic screening properties of a thin-film superconductor $\textendash$ a technique which in contrast to vortex imaging can be applied to both type-I and type-II superconductors. 

\begin{figure}
\includegraphics[width=8.6cm]{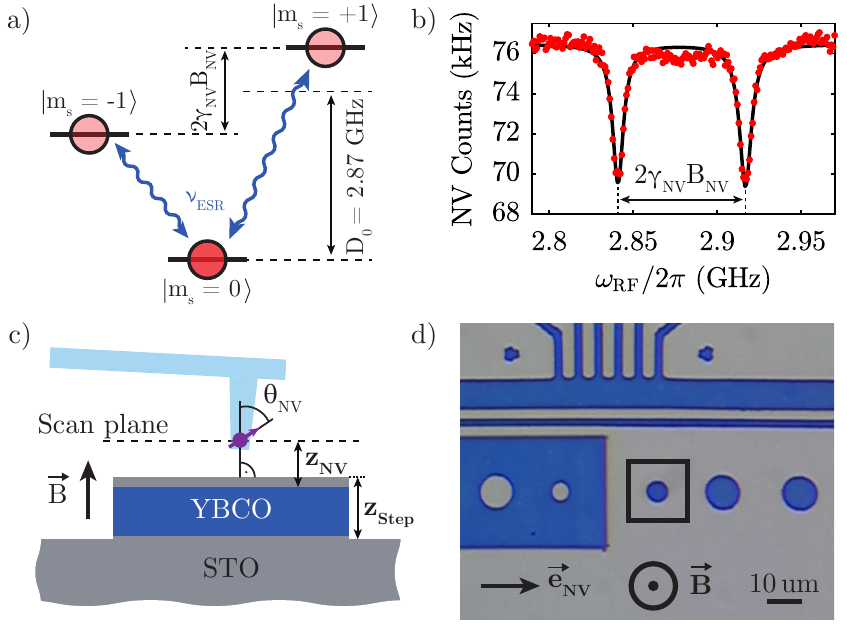}
\caption{\label{Schematics} (a) Nitrogen-Vacancy (NV) center ground state spin levels with zero-field frequency splitting $D_0$ and Zeeman splitting $2 \gamma_{\rm NV} B_{\rm NV}$ (with $\gamma_{\rm NV}=28~$GHz/T and $B_{\rm NV}$ the magnetic field along the NV axis). As indicated by the red circles, the $\ket{m_s = 0}$ spin sub-level exhibits a higher fluorescence rate than $\ket{m_s = \pm 1}$. (b) Optically detected electron spin resonance (ESR) of a single scanning NV center. (c) Schematic cross-section of the sample and the scanning probe hosting the NV center. The NV is stabilised at a distance $z_{\rm NV}$ from the superconductor surface using atomic-force distance control. (d) Top view of the micro-structured YBCO sample. Blue (grey) represents regions of YBCO (substrate), respectively. The highlighted disk with $6~\mu$m diameter is used here to study Meissner screening and a nearby four-point bridge to determine sample resistance.}
\end{figure}

\section{Experiment}
Our experiments exploit the electronic spin degrees of freedom of a single NV center in diamond  $\textendash$ a lattice point defect formed by a Nitrogen atom  adjacent to a lattice vacancy $\textendash$ as a nanoscale magnetometer. 
The NV center orbital ground state forms an electronic spin triplet consisting of the magnetic sublevels $\ket{m_s = 0}$ and $\ket{m_s = \pm 1}$ (Fig.\,\ref{Schematics}(a)), where $m_s$ denotes the magnetic quantum number along the Nitrogen-Vacancy binding axis.  
At zero magnetic field, the states $\ket{m_s = \pm 1}$ are degenerate and split by a frequency $D_0=2.87~$GHz from $\ket{m_s = 0}$.
The application of a magnetic field $B_{\rm NV}$ along the NV spin quantisation axis induces a Zeeman splitting $2 \gamma_{\rm NV} B_{\rm NV}$ of the states $\ket{m_s = \pm 1}$, where $\gamma_{\rm NV}=28~$GHz/T is the gyromagnetic ratio. Conversely, transverse fields couple to the spin degree of freedom only to second order and will be neglected here. 
The NV spin can be initialised and read out optically, since optical excitation with $532~$nm light results in spin-dependent fluorescence rates (as indicated in Fig.\,\ref{Schematics}(a)) and spin-pumping into $\ket{m_s = 0}$\,\cite{Gruber1997}. 
These combined properties enable optical detection of electron spin resonance (ESR) of NV centers, where after initialisation into the (bright) $\ket{m_s=0}$ state, a microwave driving field resonant with either of the $\ket{m_s=0}\rightarrow\ket{m_s=\pm1}$ transitions populates the less fluorescent $\ket{m_s=\pm1}$ states, resulting in a significant drop in NV fluorescence (Fig.\,\ref{Schematics}(b)). 
Such optically detected ESR thus yields a direct measure of $B_{\rm NV}$, i.e. the magnetic field projection onto the NV axis.

In order to measure  stray magnetic fields above the sample surface, we employ a single NV center located in the tip of a single-crystalline diamond scanning probe\,\cite{Appel2016}. 
This approach yields optimised sensing performance, maximal robustness and NV centers which are oriented at an angle of 54.7$^\circ$ with respect to the scanning probe (set by the crystal orientation of the tips), corresponding to an angle $\theta_{\rm NV} \approx 55^\circ$ with respect to the sample normal (Fig.\,\ref{Schematics}(c)). 
Atomic force microscopy (AFM) feedback is used to stabilise the NV center above the sample surface at typical distances $\sim50~$nm, while tip and sample are positioned and scanned using piezo actuators. 
The NV spin is addressed optically with a confocal microscope and manipulated with microwave magnetic fields generated by a close-by bonding wire spanned across the sample. 
The experiment is located in a low-vibration Helium bath cryostat with a base temperature of $4.2~$K, equipped with a superconducting vector magnet reaching fields up to $0.5~$T. 
To investigate the magnetic response of our sample, we apply an external bias magnetic field $B_z$ of $1.7~$mT perpendicular to the sample surface after cooling the superconducting film through its critical temperature $T_c$ in zero field. The continuous-wave microwave driving we employ leads to an elevated temperature $\sim8~$K (as determined by a nearby resistive thermometer) $\textendash$ a heating-mechanism which does not affect our results obtained on a high-$T_c$ superconductor and which could be readily alleviated by employing pulsed methods for ESR driving\,\cite{Dreau2011} in case of more fragile superconducting samples. 

The sample under study is a thin film of the type-II superconductor YBa$_2$Cu$_3$O$_{7-\delta}$ (YBCO)\,\cite{Wu1987}, patterned into lateral structures using optical lithography and Ar ion milling (Fig.\,\ref{Schematics}(d)). The c-axis-oriented YBCO film was grown to a thickness $d_{\rm YBCO}$ of $\sim119~$nm on top of a (001)-oriented single-crystal SrTiO$_3$ (STO) substrate using pulsed laser deposition and was covered with $\sim16~$nm of STO to avoid oxygen diffusion out of the sample\,\cite{Scharinger2012}. The film shows a $T_c\approx90~$K as measured in-situ by a four-probe conductance measurement. The data presented in this work were taken on the $6~\mu$m diameter disk highlighted in Fig.\,\ref{Schematics}(d).

\section{Results and Discussion}

\begin{figure}
\includegraphics[width=8.6cm]{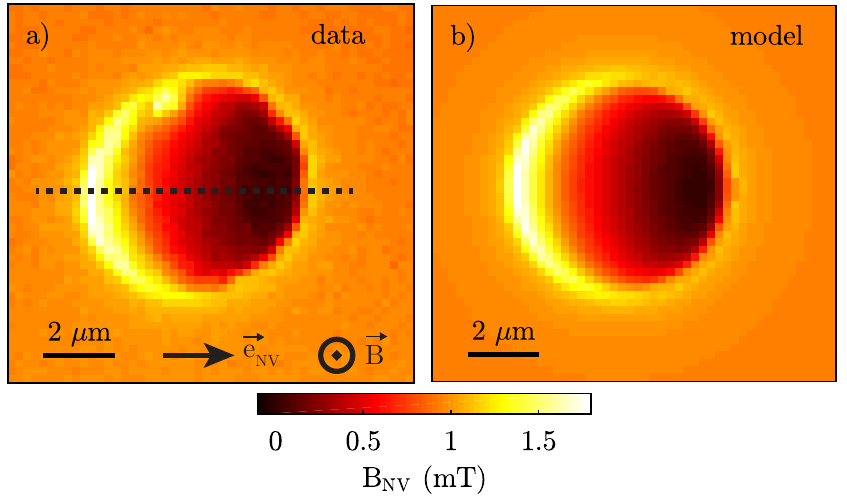}
\caption{\label{2Dmap} (a) Quantitative map of the magnetic field $B_{\rm NV}$, measured with the scanning NV spin above the YBCO disk in an external magnetic field of $1.7~$mT applied perpendicular to the sample. Low magnetic fields are observed in the center of the disk due to Meissner screening in the superconductor and maximal fields at the edges of the disk due to compression of the field lines expelled from the disk. The observed absence of rotational symmetry of $B_{\rm NV}$ around the disk center is a result of the NV orientation being away from the sample normal. The data were acquired with a pixel dwell-time of $12~$s resulting in a scan time of $8~$h for the entire scan. The green readout laser was set to a power of $350~\mu$W with a microwave power of $\sim15~$dBm sent into the cryostat. The dashed line indicates the position of the line scan in Fig.\,\ref{LineFit}. (b) Calculation of $B_{\rm NV}$ using the numerical model described in the text, with 
$\lambda=250~$nm and $z_{\rm NV}=100~$nm as manually set input parameters.}
\end{figure}

To study the nanoscale magnetic response of our sample, we first measure a full, two-dimensional stray magnetic field map of $B_{\rm NV}$ above the $6~\mu$m diameter YBCO disk (Fig.\,\ref{2Dmap}\,(a)). Magnetic field maps are created by measuring the entire ESR spectrum at each point of the scan and by determining $B_{\rm NV}$ in a subsequent fit. Our data show a strong reduction of magnetic field above the disk, while an enhanced field is measured close to the edges of the disk. This observation is in qualitative accordance with Meissner screening, where circular supercurrents inside the superconductor lead to a magnetisation opposing the external magnetic field and therefore to a reduced field in the disk center and a compression of field lines towards the disk edge. The orientation of the NV spin quantisation axis with respect to the sample normal then results in the $B_{\rm NV}$ map not being rotationally symmetric around the disk center. 

In order to gain further understanding of these observations, we numerically calculate the current density $j_s(r,z,\lambda)$ induced in a superconducting disk exposed to an external perpendicular magnetic field. Due to the cylindrical symmetry of our sample, it is sufficient to only consider the $z$-component of the second London equation in cylindrical coordinates\,\cite{Tinkham2004}
\begin{equation}
\label{2ndLondon}
-\frac{B_z}{\mu_0\lambda^2} = (\nabla\times \vec{j}_s)_z = \frac{(\vec{j}_s)_\phi}{r} + \frac{\partial (\vec{j}_s)_\phi} {\partial r},
\end{equation}
where the subscripts indicate the vectorial components in cylindrical coordinates and $\mu_0 = 4\pi \times 10^{-7}~$H/m is the vacuum permeability. Since only the supercurrent density in azimuthal direction $(\vec{j}_s)_\phi$ appears in this equation, we denote this quantity as $j_s$ in the following. We find $j_s$ by numerically solving Eq.\,(\ref{2ndLondon}) in a disk with a diameter of $6~\mu$m and a thickness of $119~$nm using a grid of $300\times12$ elements, corresponding to a resolution of $10~$nm. As boundary condition, we set $j_s(r=0) = 0$ in the center of the disk and determine $j_s(r,z)$ with the Euler method. Our calculation of $j_s$ converges by iteratively calculating $j_s(r,z)$, updating $B_z(r,z)$ and then recalculating $j_s(r,z)$. The magnetic field $B_{\rm NV}$ is then calculated at the position of the sensor using Biot-Savart's law, taking into account the topography of the disk.
Our procedure yields good qualitative agreement with our data when using a penetration depth $\lambda=250~$nm determined earlier\,\cite{Thiel2016} and an estimated NV-to-superconductor distance $z_{\rm NV}=100~$nm (Fig.\,\ref{2Dmap}(b)). This agreement motivates further quantitative studies for an independent determination of $\lambda$ from our data, as discussed in the following.

\begin{figure}
\includegraphics[width=8.6cm]{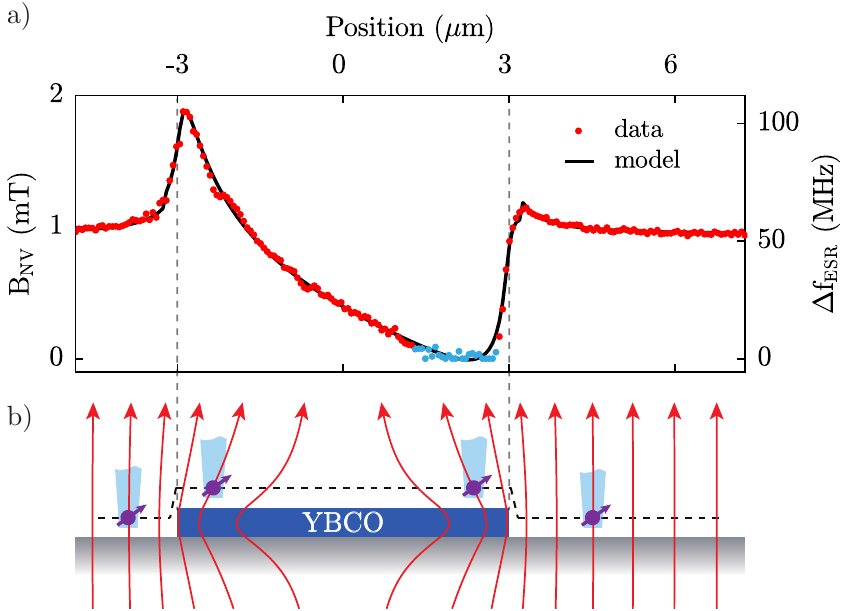}
\caption{\label{LineFit} (a) Measurement of $B_{\rm NV}$ across the YBCO disk along the trajectory indicated in Fig.\,\ref{2Dmap}\,(a). The black line in (a) shows the fit with the numerical model which yields a penetration depth of $\lambda=249\pm5~$nm. Data points marked in blue were excluded from the fit (see text). The data integration time was $24~$s per point resulting in $80~$min for the entire scan. A laser power of $470~\mu$W and a microwave power of $\sim15~$dBm were applied to the NV spin. (b) Magnetic field lines around the YBCO disk in the Meissner state, along with a sketch of the experimental setup.  The field lines together with the NV spin quantisation direction (purple arrows) highlighted for various positions illustrate the asymmetry in $B_{\rm NV}$ observed in the data. The dashed line illustrates the topography of the scan which is taken into account in the calculation of $B_{\rm NV}$.}
\end{figure}

In order to quantitatively analyse our data and determine the penetration depth $\lambda$, we perform a high resolution line scan of $B_{\rm NV}$ across the YBCO disk, along the trajectory indicated in Fig.\,\ref{2Dmap}\,(a). The resulting data (Fig.\,\ref{LineFit}) show the same global features already discussed for the 2D map, all while providing greater detail and a higher sampling rate for subsequent, quantitative fitting using the same method as described above. 
For the few data points (marked in blue in Fig.\,\ref{LineFit}) where the observed NV ESR splitting falls short of the ESR linewidth of $\sim8~$MHz, our determination of $B_{\rm NV}$ is unreliable due to poor fitting quality and the effect of local strain on the NV spin splitting\,\cite{Doherty2013}. We therefore excluded these points in the subsequent analysis.
From a fit to these data, we seek to extract the three key unknown parameters $\lambda$, $z_{\rm NV}$, and $\theta_{\rm NV}$. 
To obtain the best estimate for their values and respective errors, we include in our model three additional nuisance parameters, namely the lateral position of the superconductor within the scan, the magnitude of the externally applied field, and a calibration factor to scale the $x$-axis. 
These nuisance parameters are required for the model and the error estimation but carry little information about the final values of $\lambda$, $z_{\rm NV}$, and $\theta_{\rm NV}$. 
The resulting model fit (Fig.\,\ref{LineFit}) shows excellent agreement with data and yields $\lambda=249\pm5~$nm, $z_{\rm NV}=70\pm10~$nm and $\theta_{\rm NV}=55.3\pm0.7^\circ$. 
Here, the error bars denote $2 \sigma$ confidence intervals, which we calculate from the diagonal elements of the covariance matrix of the free parameters, i.e. the width of the marginal probability density function for each parameter. The error bars therefore in principle account for correlations between all free parameters. We find, however, significant correlations only between $\lambda$ and $z_{\rm NV}$.
The value of $\lambda$ we find with this approach agrees well with a previous and independent measurement we performed in the same material\,\cite{Thiel2016}, which gives further confidence in the validity of our approach. Taking into account the $16~$nm thick STO capping layer, we reach a NV-to-sample distance of $\sim50~$nm, which constitutes an important benchmark for the resolution capability of our setup. 

\begin{figure}
\includegraphics[width=8.6cm]{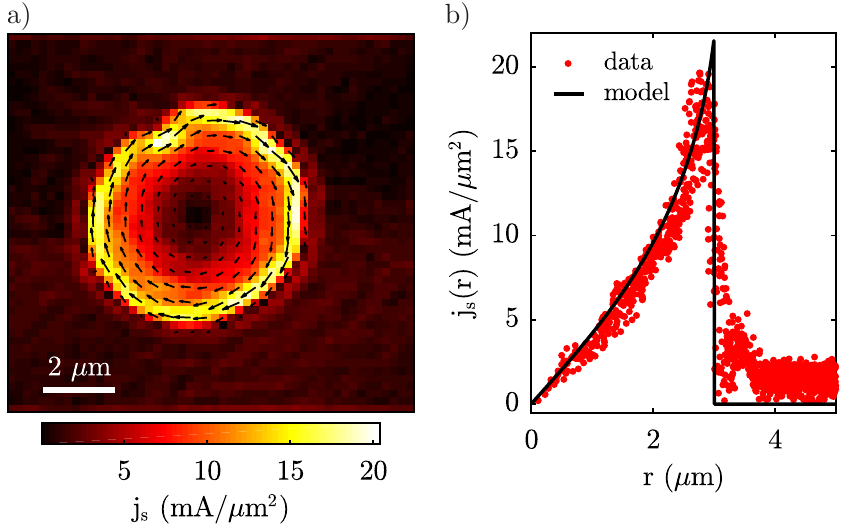}
\caption{\label{CurrentDensity} (a) Current density $j_s$ reconstructed by magnetic field reverse-propagation from the data in Fig.\,\ref{2Dmap}\,(a). The circular currents generate a magnetic field counteracting the external magnetic field. (b) Azimuthal average of the current density $j_s(r)$ as a function of distance to the disk center, along with the current density calculated in the numerical model.}
\end{figure}

Lastly, we take advantage of the one-to-one mapping between a two-dimensional current distribution and a magnetic field map in a plane away from the field source and use the 2D data presented in Fig.\,\ref{2Dmap}\,(a) to determine the supercurrent distribution in the YBCO disk which is responsible for Meissner screening at first place. To that end, we employ a well-established reverse-propagation method\,\cite{Roth1989} to determine $j_s$ in the superconductor. In short, the current density is obtained by applying the inverse Biot-Savart law to the measured Oersted magnetic field in Fourier space. The method assumes no $z$-dependence of the current distribution inside the disk, which, however, is a valid approximation in our case since $d_{\rm YBCO} \ll \lambda$. This reverse propagation process has a tendency to exponentially amplify high-frequency components, and therefore noise, in the imaging data. To counteract this detrimental tendency, it is common practice\,\cite{Roth1989} to filter out these high-frequency components using a low-pass filter with a typical cutoff frequency corresponding to the inverse sample-to-sensor distance $z_{\rm NV}$. In our case, the sampling distance in the image was comparable to $z_{\rm NV}$ and therefore provided a suitable filter on its own and no further numerical filtering was performed.
The resulting reconstructed current density shows the expected circular screening currents inside the YBCO disk (Fig.\,\ref{CurrentDensity}). Moreover, the current reconstruction reveals the position and shape of a region in the sample where superconductivity is suppressed and the current therefore deviates from the circular shape of the disk. Signatures of this were already visible in Fig.\,\ref{2Dmap}\,(a), but we now visualise it more directly through the impact it has on the current flow in the superconductor. Note that no hint of a defect was visible in the topography of the sample. While further studies are required to assess the origin of this feature in $j_s$, we speculate that it may either constitute a vortex trapped at the edge of the YBCO disc or a defect connected to material impurities introduced during growth or micro-patterning.

For the circulalry symmetric part of the current distribution, we can compare $j_s$ as obtained from reverse-propagation to the one calculated by our numerical model described above. For this, we determine the azimuthal average of the reconstructed map $j_s(r)$, while excluding currents close to the irregularity discussed in the previous paragraph (Fig.\,\ref{CurrentDensity}(b)). Close to the center of the disk, we observe a linear increase of $j_s(r)$ with $r$, corresponding to a constant curl of $\vec{j}_s$ due to a homogeneous field $B_z > 0$ (c.f. Eq.\,(\ref{2ndLondon})). Conversely, the supercurrent increases super-linearly within a distance $\Lambda$ from the edge of the disk, where $\Lambda = 2 \lambda^2 / d_{\rm YBCO} \approx 1.04~\mu$m is the effective penetration depth in a superconductor with thickness $d_{\rm YBCO}$\,\cite{Tinkham2004}.

\section{Conclusion and summary}

In conclusion, we have experimentally demonstrated the use of scanning NV magnetometry to address nanoscale properties of superconductors through stray magnetic field imaging. Specifically, our experiments performed on micron-scale YBCO disks allowed us to quantitatively determine the local value of the London penetration depth in the material and to assess the impact of disorder in our thin superconducting film through supercurrent reconstruction. These results establish NV magnetometry as an attractive resource to study magnetic properties of superconductors and open the way to addressing open questions in mesoscopic superconductivity. 
Specifically, our quantitative and high-resolution imaging method could be employed to study supercurrents beyond simple Meissner screening, such as presented in recent studies on superconductor-ferromagnet heterostructures\,\cite{Flokstra2018} or anomalous surface currents arising from Andreev bound states\,\cite{Iniotakis2008}, which may enable determination of the superconductors order parameter symmetry. 
Our direct access to the nanoscale magnetic response of superconductors offers attractive prospects to study the precursor phase of superconductivity at elevated temperatures above $T_c$, where island superconductivity could be observed in real space using our approach\,\cite{Kresin2006}. Furthermore, novel quantum sensing techniques recently developed for NV magnetometry offer attractive prospects in the present context. For example, noise spectroscopy\,\cite{Bar-Gill2012} and NV relaxometry\,\cite{Tetienne2013,Schmid-Lorch2015} could be used to assess magnetic flux noise in superconductors $\textendash$ a general problem that plagues superconductor-based quantum devices\,\cite{Muller2017} and that NV magnetometry could help alleviate in the future.

\section{Acknowledgements}

We thank D. Budker for fruitful discussions. We gratefully acknowledge financial support through the NCCR QSIT, a competence center funded by the Swiss NSF, through the Swiss Nanoscience Institute (SNI) and through SNF Grant Nos. 142697 and 155845. This research has been partially funded by the European Commission’s 7. Framework Program (FP7/ 2007-2013) under Grant Agreement No. 611143 (DIADEMS). We also acknowledge support by the COST action NANOCOHYBRI (CA16218).


%

\end{document}